\documentclass{ceab}   

\usepackage{epsfig}     
\usepackage{graphicx}   

\usepackage{ceabbib}     
\usepackage[T1]{fontenc}

\usepackage{multicol}
\def\farcs{\hbox{$.\!\!^{\prime\prime}$}}
\def\farcd{\hbox{$.\!\!^{\circ}$}}

\def\gore{MUDEHaR: magnetic O-type stars}

\usepackage{siunitx}

\begin{document}

\title{Multi-epoch precise photometry from the ground: MUDEHaR, magnetic stars and everything around}

\author{G. HOLGADO$^{1}$, J. MAÍZ APELLÁNIZ$^2$ and 
J.~A. CABALLERO$^{2}$,\\ + the MUDEHaR team
\vspace{2mm}\\
\it $^1$Instituto de Astrofísica de Canarias, \\ 
\it Avenida Vía Láctea, E-38205 La Laguna, Tenerife, Spain\\
\it $^2$Centro de Astrobiología (CAB), CSIC-INTA,\\ 
\it Campus ESAC, C. Bajo del Castillo s/n, E-28692 Madrid, Spain
}

\maketitle

\begin{abstract}
MUDEHaR is an on-going multi-epoch photometric survey with two narrow filters in H$\alpha$ and the calcium triplet window that uses the T80Cam wide-field imager at the JAST/T80 telescope at Spanish Javalambre astronomical observatory. It is obtaining 100 epochs/year per field for 20 fields in the Galactic disk, each of 2\,deg$^2$, for a total of 40\,deg$^2$. Focused on stellar clusters and HII regions including bright stars, its main objective is to detect tens of thousands OB stars that present emission variability in H$\alpha$ on days-months-years scale. The observed targets include magnetic massive stars, pulsating stars, and all kinds of variable stars.
Among our driving scientific objectives of MUDEHaR observations is to identify potential magnetic candidates in massive stars. Only 10--20\,\% of OB stars display a measurable magnetic field, and its origin is still in debate. We outline here the multi-step process involved in identifying OB magnetic stars, highlighting the significance of MUDEHaR observations in this process.
\end{abstract}


\section{Introduction}



MUDEHaR is a photometric survey focused on finding stars with variability in H$\alpha$, in a region of the sky of 40\,deg$^2$ using the T80Cam camera installed at the JAST/T80 telescope of the Javalambre astronomical observatory, situated in Teruel, Spain. The study includes 100 epochs per year, for each of the 20 fields of 2\,deg$^2$ each distributed throughout the Galactic disk.
The main objective is to focus on the massive stars that present emission in H$\alpha$, which are peculiar stars with magnetic fields or disk emission (Oe and Be), Wolf-Rayet stars, central stars of planetary nebula, eclipsing binaries, ellipsoidal variables , and pulsating stars. In this way, it is intended to provide accurate statistics on these type of interesting objects when it comes to understanding the formation and evolution of massive stars. A second filter centered on the Ca II near-infrared triplet allows us to maximize source detection, and accurately link to 2MASS and \textit{Gaia} photometry and astrometry.

Historically, there have been other large-area or whole-sky ongoing photometric surveys including a precise measurement of H$\alpha$ (EGAPS, VPHAS +, J-PLUS, S-PLUS, J-PAS, OmegaWhite, HAWKs, ADHOC, HOYS, STREGA , PTF/iPTF, GALANTE). However, among them only two present multi-epoch capabilities: OmegaWhite \citep[using the wide-field instrument OmegaCAM,][with the 2.65m VLT Survey Telescope at Paranal Observatory]{Kupfer2017} in search of short period variables, and PTF/ iPTF \citep[Palomar Transient Factory/ intermediate PTF,][with 48 inch Samuel Oschin
Telescope ]{Yu2018} with its recent inclusion of a pair of new narrow band filters on and off the H$\alpha$ line in some clusters of interest.

A first sample of interest to be studied are the magnetic OB stars. 
Making up only 10\% of OB stars \citep[MiMeS collaboration --][]{Wade2016, Negueruela2004}, they constitute a minority sample, and the question of their origin is a matter of debate.
Some authors proposed an origin from stellar merger phenomena \citep{Schneider2019}, others study the “fossil field” scenario, from pre-main sequence evolution \citep{Cowling1945}, but based on its percentage of the magnetic vs. total sample of OB stars the origin of magnetic fields in OB stars is still under debate. 
Another batch of interesting targets for MUDEHaR scientific team are the Oe and Be emission stars. Their variability is proposed to be linked to episodic mass ejections and a circumstellar disk, often induced by a combination of fast rotation and stellar pulsations \citep[][and references therein]{Harmanec1983,Hubert1997,Kourniotis2014}.
In addition, although the general mapping is oriented to OB stars, the observations include tens of thousands of stars of other spectral types and could represent an interest results in case of detecting variability in H$\alpha$ and the near-infrared continuum (e.g. Herbig Ae/Be and T Tauri stars). With a rough estimate of between 40 000 and 100 000 sources per field, just 1\% variable stars would imply around 400 targets per field.

\section{Observations}

MUDEHaR unique characteristics incude precise H$\alpha$ variability measures, through narrow filters, on the scale of days-months-years, for tens of thousands of stars (from AB mag = 3 down to 17 with S/N between 70 and 500). The small pixel size (0.55 arcsec/pix) naturally overcomes the crowding and nebular problems that affect other surveys such as \textit{Gaia} or TESS. Our proposal focuses on stellar clusters and HII regions including bright stars, as opposed to the many surveys focused away from the disk to avoid extinction. It has a dynamic range of magnitudes that extends much more towards the bright stars than the rest of the deep-focused surveys in the literature, allowing comparison with photometry that has been used as a reference for a long time such as Hipparcos \citep{Perryman1998} .

\subsubsection*{Filters}

MUDEHaR uses two narrow filters selected according to the following criteria:
\begin{itemize}
	\item J0660 is a narrow-band filter situated on top of the H$\alpha$ line. It is capable of precise measurements down to several millimagnitudes according to previous results. H$\alpha$ is an indicative of stellar activity, and measurements centered on this spectral line aid in detecting emission-line stars and assessing nebular contamination.
	\item J0861 is an intermediate filter in the calcium triplet window. It is being used to detect the largest possible number of stars to tie the astrometry and photometry with 2MASS. Capable of precise measurements down to few millimagnitudes according to previous results. It present synergies with the \textit{Gaia}-RVS information as the \textit{Gaia} consortium intends to provide pseudo-photometry derived from RVS spectroscopy.
\end{itemize}

\subsubsection*{Exposure time}
The acquisitions are derived from four distinct exposure durations.
The final result would be 100 epochs per year for each field (i.e. each star).
\begin{itemize}
	\item 10 exposures of 50 s each (“long” exposures, for 13.0--18.0\,mag) at two different air masses (5 in each). The fields are observed at different air masses (high/low $z$) to provide a more accurate photometric calibration and eliminate possible color effects introduced by the atmosphere each night. 
	\item 5 exposures of 10 s each (“intermediate” exposures, for 11.0--13.0\,mag).
	\item 5 exposure of 1 s each (“short” exposure, for 8.5--11.0\,mag).
	\item 5 exposures of 0.1 s each ("very short" exposures, for 6.0--8.5\,mag).
\end{itemize}

Each field is planned to be visited a maximum of 20 times throughout the year, capturing a minimum of 5 epochs on each occasion. This enables the assessment of variability on an hourly scale. Subsequently, these pointings are repeated during the months when each field is visible, with a gap of at least one week between repetitions.

\subsubsection*{Layout}

\begin{figure}
	\includegraphics[scale=0.35]{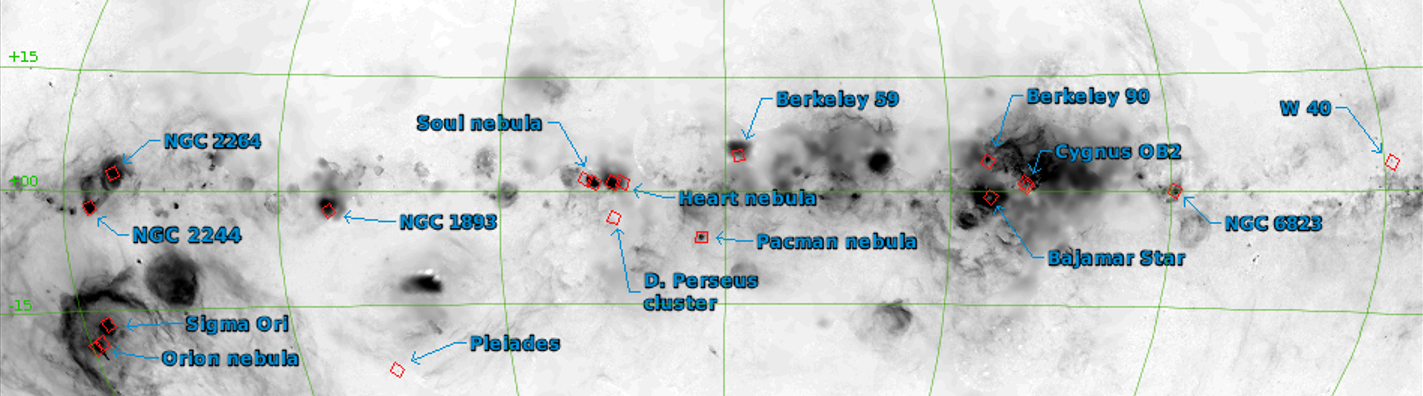}%
	\caption{ Footprint of the 20 MUDEHaR fields (in red). The background is an H$\alpha$ image of \cite{Finkbeiner2003} in Galactic coordinates using an Aitoff projection and logarithmic scale.}
	\label{figuraMW}
\end{figure}

Our survey focus on regions of the Galactic disk, which are often removed from other surveys due to the intrinsic difficulties involved. In total there are 20 regions of 2\,deg$^2$ each. The selection is made prioritizing areas of recent star formation and HII regions, low extinction, and homogeneous distribution during the observation semesters. 17 of the fields are directly over the Galactic disk, and 3 fields are in the Orion star-forming zone, see Fig.~\ref{figuraMW}. We emphasize again that the main crowding difficulties that these areas present are not a problem for the T80CAM and its powerful pixel-scale (0.55 arcsec/pix).
For a quick and simple check to see if any object of interest is in MUDEHaR, there is a tool called InMUDEHaR (https://github.com/gholgadoa/InMUDEHaR). You provide a list of coordinates, and it informs you whether they will be observed by MUDEHaR in any of the 20 fields.

As a final product, a catalog of photometric values in H$\alpha$ and near-infrared will be made available to the scientific community for all sources visible in the optical for each of the field/night observed. 
The preliminary results reveal a range of 17 000 to 120 000 valid sources per field. Instances where the measurement is deemed invalid (due to issues like saturation, contamination, or faintness beyond the suitable exposure time) are marked as U (Undone) and with a corresponding value of 99.99. For cases where the measurement is valid but the exposure time is not optimal for the star's magnitude, a value is calculated, albeit with a substantial uncertainty. The following section outlines the procedure adopted to address these scenarios.
This catalog of variable stars will represent a valuable support for satellite missions covering other regions of the sky, or similar zones with a different emphasis on the multiepoch aspect. 

Finally, the combination of high-quality images of different exposure times allows generating an arcsecond-resolution continuum-subtracted H$\alpha$ map of the northern Galactic plane with spatial resolution much better than previous studies, and will be very useful to study variability in HII regions, planetary nebulae, the movement of runaway and walkaway stars, and the diffuse H$\alpha$ emission.

\section{Results}

The initial results from the first two months of observations if MUDEHaR (January and February 2023) have been successfully processed. The processing encountered no issues as the pipeline had been previously generated and validated for a similar project with the same type of data.
As anticipated, these observations from the first MUDEHaR campaign revealed a broad dynamic range in magnitudes, encompassing bright and quite dim stars (AB mag 3-17 with S/N > 100), and confirmed the powerful spatial resolution capability on the sky (0.55 arcsec/pix).
We compiled a mean of 10 measurements (i.e. epochs) for $\sim$75~000 individual stars in each of the 12\,deg$^2$ investigated to that date.

\subsubsection*{Comparison with \textit{Gaia}}

\begin{figure}
	\centering
	\includegraphics[scale=0.5]{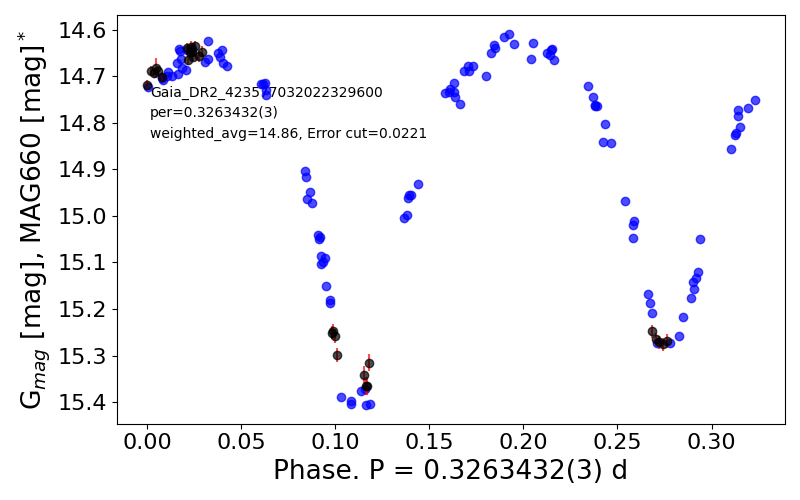}%
	\caption{Folded light curve of \textit{Gaia} $G$ magnitude data for the system indicated in blue, with overplotted MUDEHaR data in black (uncertainties in red). \textit{Gaia} period from \cite{Mowlavi2023}. The period, mean magnitude, and error limit (see text), are indicated in the figure.}
	\label{MH1_example}
\end{figure}

During the project development, we anticipated that MUDEHaR measurements would be as precise as the available \textit{Gaia} epoch photometry. Therefore, one of our initial validations involved identifying known eclipsing binaries among the stars in the observed MUDEHaR fields during this first run and confirming that MUDEHaR measurements could accurately fit the light curve constructed with \textit{Gaia} data for those systems. The results were highly satisfactory, with MUDEHaR points aligning almost perfectly with the \textit{Gaia} curves (see Fig.~\ref{MH1_example}), demonstrating remarkably competitive uncertainty for an 80\,cm diameter telescope.

The comparisons indicate accuracy and precision levels in line with the expectations outlined in the proposal, comparable to \textit{Gaia} epoch photometry. We successfully fitted \textit{Gaia} curves with MUDEHaR data for stars with $G$ = 15\,mag, achieving nominal expected errors of 0.01 mag for the J0660 filter and 0.005 mag in the J0861 filter. For brighter stars, the uncertainties are even lower, and for stars up to a magnitude of 17 we could still reproduce light curves perfectly, though with larger uncertainties of up to 0.05 mag.

\begin{figure}
	\centering
	\includegraphics[scale=0.7]{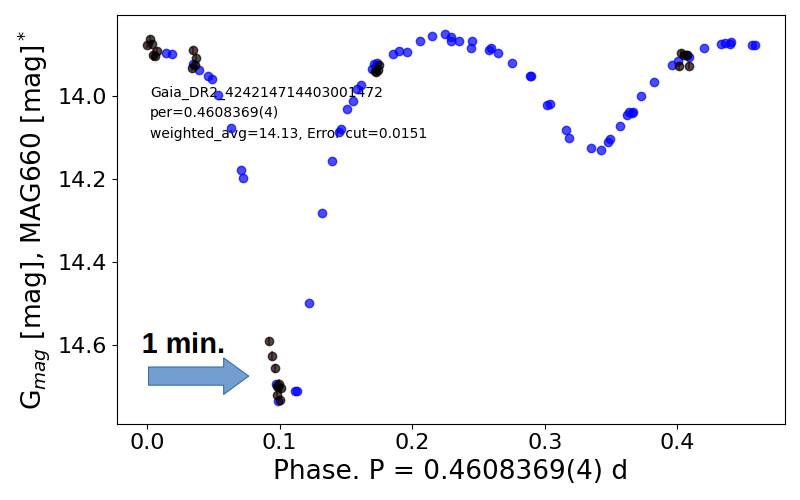}%
	\caption{Same as Fig.~\ref{MH1_example}, but for a very short period eclipsing binary system, with consecutive exposures separated by 1 minute.}
	\label{MH1_precise}
\end{figure}

It is worth noting the high reliability of measurements in terms of accuracy and precision. For a short-period eclipsing star ($\sim$0.4 d), measurements separated by just minutes could accurately depict the descent in magnitude of an eclipse (see Fig~\ref{MH1_precise}).

\subsubsection*{Magnetism and photometric variability}

\begin{figure}
	\centering
	\includegraphics[width=\textwidth]{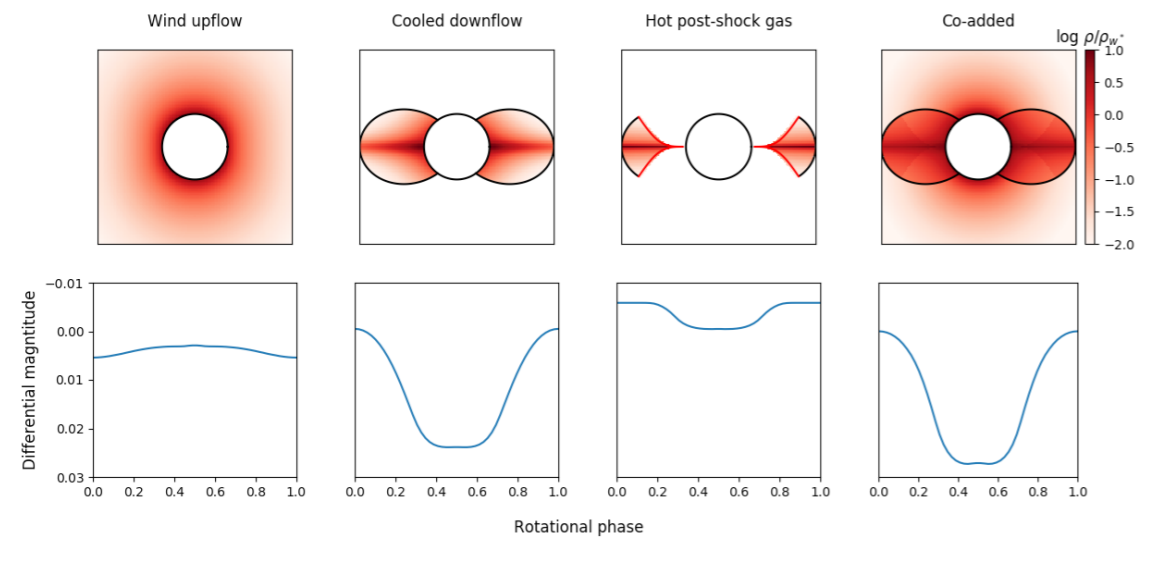}%
	\caption{Photometric variability due to magnetic structure. \textit{Top row}:
		Density structure for different components in the Analytical Dynamical Magnetosphere
		(ADM) models, and the resulting co-added final structure. \textit{Bottom row}: Synthetic light curve
		corresponding to the different ADM components and the resulting light curve from the coadded density. Variations of up to 0.03\,magnitudes are simulated. \cite{Munoz2020}}
	\label{Munoz2020}
\end{figure}

\begin{figure}
	\includegraphics[width=0.5\textwidth]{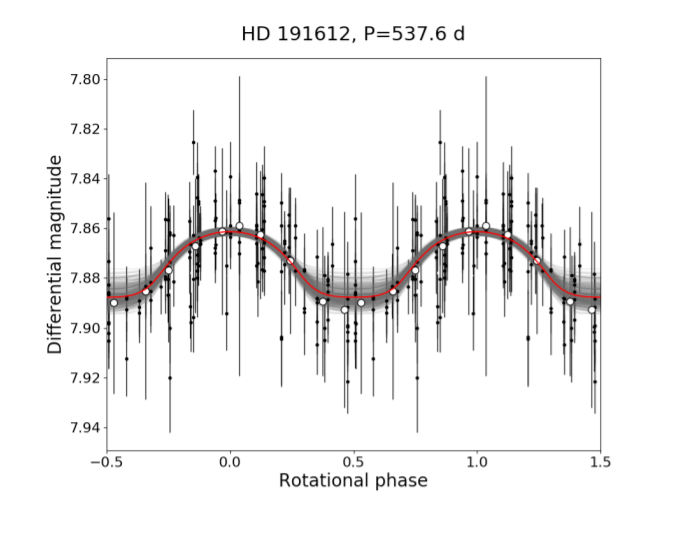}%
	\includegraphics[width=0.5\textwidth]{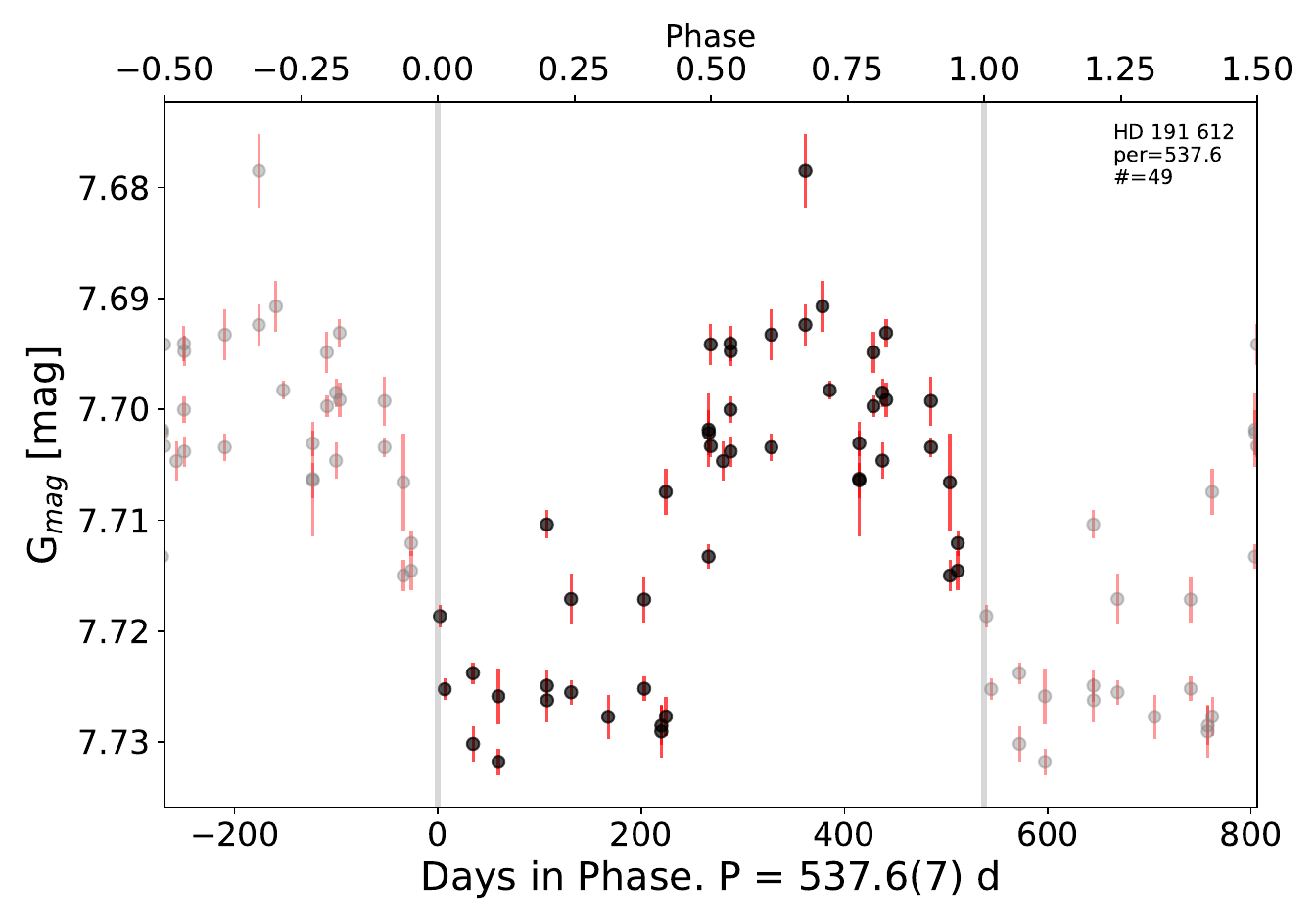}%
	\caption{ \textit{Left:} Hipparcos light curve for a known magnetic O-type star (HD\,\num{191612}) with errors. The curve of best-fit ADM is overplotted in red. \textit{Right:} Folded light curve of Gaia $G$ magnitude data for the system indicated. Observations (the dots) include uncertainties. In grey are repeated points beyond the central period. The period and number of points are indicated in the figure. \cite{Munoz2020}}
	\label{DavidUraz2021}
\end{figure}

There has been an effort to model the photometric variability produced by the presence of magnetic fields \citep{Munoz2020}. Theoretical models show that the shape of the variability curve (photometric variability for broad filters including H$\alpha$) could depend on the wind structure associated to the magnetic field (Fig.~\ref{Munoz2020}), and observations of well-known magnetic stars hint towards its validation (Fig.~\ref{DavidUraz2021}). As part of the MOBSTER collaboration series of paper, \cite{David-Uraz2021} presented that photometric variability in TESS is a good indicator in the search of magnetic B-type stars. In addition, several members of the MUDEHaR team participated in the identification of magnetic O-type stars through rotational modulation of TESS photometry. We must note that TESS pixels are relatively large ($21\times21$\,arcsec$^2$), and so targets may include flux from multiple nearby stars, something that MUDEHaR could particularly help with.
We have confirmed that these variations in amplitude and periodicity are detectable with MUDEHaR data. However, substantial detections may require a considerable amount of additional data, beyond the initial observations from these first months.

\section{Conclusions and future work}

The main features of MUDEHaR are summarized in Table~\ref{tab1}. The observations are already underway and will last for five years. The calibration results are ideal, meeting the requirement of being comparable to \textit{Gaia} photometry.
The final result will be catalogs with information on thousands of variable stars, of all types, along with sky-mapping of H$\alpha$ and near-infrared variability.

\begin{table}
	\hspace{-2.0cm} 
	\fontsize{9}{11}\selectfont
	\begin{tabular}{lll}
		{\it Detector:}        & \multicolumn{2}{l}{1\farcd4$\times$1\farcd4 continuous FOV with 0\farcs55 pixels.} \\
		{\it Footprint:}       & \multicolumn{2}{l}{17 fields in the Galactic north plane: $\;\,|b| < 3^{\rm o}$ + $|\delta| > 0^{\rm o}$.} \\
		& \multicolumn{2}{l}{3 fields in the Orion region; total of $\sim$40 sq. dg., Fig.~\ref{figuraMW}.} \\
		{\it Epochs:}    & \multicolumn{2}{l}{100 per field} \\
		{\it Exposure time total:}  & \multicolumn{2}{l}{200$\times$0.1 s (when required) + 200$\times$1 s + 200$\times$10 s + 200$\times$50 s (low z) + 200$\times$50 s (high z)} \\
		{\it Magnitude range:} & \multicolumn{2}{l}{Unsat. AB mag 3-17 with S/N $>$ 100 in both filters, detect. to AB mag 19-20.} \\
		{\it Precision threshold:}    & \multicolumn{2}{l}{J0660 (15.5 nm): 0.01 mag; J0861 (40.0 nm): 0.005 mag.} \\
		{\it Survey dates:}    & \multicolumn{2}{l}{2023-2028.} \\
		{\it Filters:}         & J0660 & H$\alpha$ line, pure nebular images + emission-line star detection. \\
		& J0861 & Ca IRT, tie-in with Gaia-RVS and 2MASS, extinction typing. \\
	\end{tabular}
	\caption{MUDEHaR in a nutshell}
	\label{tab1}     
\end{table}

The zeroth data release (DR0) will be the catalog of a specific field and will be built with the aim of being a methodology benchmark. It will be compiled from the first 100 epochs of one of the test fields for which previous processing is already complete: NGC 2244. This field is observable in the first months of the year. The DR0 will be ready by March 2024, so that any necessary adjustments for the MUDEHaR observations can be applied in the period of maximum observational pressure, between October and December. After the compilation of the 100 epochs for each field in the first year, DR1 will be constructed with 20 catalogs, one per field, as well as the mono-filter images of the epochs of the first year. The intended date of release of DR1 is September 2024, after the low pressure window of observations between March and April. After this, annual renewals can be added on these same dates. Finally, in these successive DRs, the combined bi-color images of each observation night will be made available, as well as their sequence. This process would be repeated during the five awarded years, until finally providing a total summary catalog of all the observations for each field.

In the future, this data will be employed to apply period determination methods using all available information in MUDEHaR for each object of interest. Our aim is to identify potential candidates for magnetic stars among the most massive ones.

\section*{Acknowledgements} 
Based on observations made with the JAST/T80 telescope at the
Observatorio Astrofísico de Javalambre, in Teruel, owned, managed and
operated by the Centro de Estudios de Física del Cosmos de Aragón.
We acknowledge financial support from the Agencia Estatal de
Investigaci\'on (AEI/\num{10.13039}/\num{501100011033}) of the Ministerio de
Ciencia e Innovaci\'on and the ERDF ``A way of making Europe'' through
projects
PGC2018-\num{095049}-B-C22
PID2021-\num{122397}NB-C21,
PID2022-\num{13664}0NB-C22 (grant 2022-AEP~005) and,
PID2022-\num{137241}NB-C42.
This work was also funded by the COST Action CA18104 MW-Gaia.

\newpage


\bibliographystyle{ceab}
\bibliography{sample}

\end{document}